\documentclass[apj]{emulateapj}
\usepackage{apjfonts,graphicx}

\newcommand{\kms}{\hbox{km\,s$^{-1}$}}

\newcommand{\water}{\hbox{H$_{\rm 2}$O}}
\newcommand{\watertrans}{\hbox{H$_{\rm 2}$O$~(2_{\rm 02}-1_{\rm 11}$)}}

\newcommand{\jykms}{\hbox{Jy\,km\,s$^{-1}$}}

\newcommand{\coeight}{\hbox{CO $J$=8-7}}
\newcommand{\coten}{\hbox{CO $J$=10-9}}
\newcommand{\cofive}{\hbox{CO $J$=5-4}}
\newcommand{\jeight}{\hbox{$J$=8-7}}
\newcommand{\jten}{\hbox{$J$=10-9}}
\newcommand{\jfive}{\hbox{$J$=5-4}}

\slugcomment{}

\shorttitle{ALMA Long Baseline Observations of SDP.81 at $z$=3.042}
\shortauthors{ALMA Partnership et al.}

\begin{document}

\title{ALMA Long Baseline Observations of the Strongly Lensed Submillimeter Galaxy \hbox{HATLAS J090311.6+003906} at $z$=3.042}

\author{ALMA Partnership,
C. Vlahakis\altaffilmark{1,2}, 
T. R. Hunter\altaffilmark{3},
J. A. Hodge\altaffilmark{3},
L. M. P\'erez\altaffilmark{4},
P. Andreani\altaffilmark{5},
C. L. Brogan\altaffilmark{3},
P. Cox\altaffilmark{1,2},
S. Martin\altaffilmark{6},
M. Zwaan\altaffilmark{5},
S. Matsushita\altaffilmark{7},
W. R. F. Dent\altaffilmark{1,2},
C. M. V. Impellizzeri\altaffilmark{1,3},
E. B. Fomalont\altaffilmark{1,3},
Y. Asaki \altaffilmark{8,9},
D. Barkats\altaffilmark{1,2},  
R. E. Hills\altaffilmark{10},
A. Hirota\altaffilmark{1,8},  
R. Kneissl\altaffilmark{1,2},
E. Liuzzo\altaffilmark{11},
R. Lucas\altaffilmark{12},
N. Marcelino\altaffilmark{11},
K. Nakanishi\altaffilmark{1,8},  
N. Phillips\altaffilmark{1,2}, 
A. M. S. Richards\altaffilmark{13},
I. Toledo\altaffilmark{1},
R. Aladro\altaffilmark{2},
D. Broguiere\altaffilmark{6}, 
J. R. Cortes\altaffilmark{1,3},
P. C. Cortes\altaffilmark{1,3},
D. Espada\altaffilmark{1,8},
F. Galarza\altaffilmark{1},
D. Garcia-Appadoo\altaffilmark{1,2}, 
L. Guzman-Ramirez\altaffilmark{2},  
A. S. Hales\altaffilmark{1,3},
E. M. Humphreys\altaffilmark{5},  
T. Jung\altaffilmark{14},   
S. Kameno\altaffilmark{1,8}, 
R. A. Laing\altaffilmark{5},     
S. Leon\altaffilmark{1,2},
G. Marconi\altaffilmark{1,2}, 
A. Mignano\altaffilmark{11},
B. Nikolic\altaffilmark{10},
L. -A. Nyman\altaffilmark{1,2}, 
M. Radiszcz\altaffilmark{1}, 
A. Remijan\altaffilmark{1,3},
J. A. Rod\'on\altaffilmark{2},  
T. Sawada\altaffilmark{1,8},
S. Takahashi\altaffilmark{1,8},
R. P. J. Tilanus\altaffilmark{15},   
B. Vila Vilaro\altaffilmark{1,2}, 
L. C. Watson\altaffilmark{2},
T. Wiklind\altaffilmark{1,2},
Y. Ao\altaffilmark{8},
J. Di Francesco\altaffilmark{16},
B. Hatsukade\altaffilmark{8},
E. Hatziminaoglou\altaffilmark{5},
J. Mangum\altaffilmark{3},
Y. Matsuda\altaffilmark{8},
E. van Kampen\altaffilmark{5},   
A. Wootten\altaffilmark{3}, 
I. de Gregorio-Monsalvo\altaffilmark{1,2},
G. Dumas\altaffilmark{6},
H. Francke\altaffilmark{1},
J. Gallardo\altaffilmark{1},
J. Garcia\altaffilmark{1},
S. Gonzalez\altaffilmark{1}, 
T. Hill\altaffilmark{1,2},
D. Iono\altaffilmark{8},
T. Kaminski\altaffilmark{2}, 
A. Karim\altaffilmark{17},
M. Krips\altaffilmark{6},
Y. Kurono\altaffilmark{1,8},
C. Lonsdale\altaffilmark{3},
C. Lopez\altaffilmark{1},
F. Morales\altaffilmark{1},
K. Plarre\altaffilmark{1},
L. Videla\altaffilmark{1},
E. Villard\altaffilmark{1,2},
J. E. Hibbard\altaffilmark{3},
K. Tatematsu\altaffilmark{8}
}

% JAO
\altaffiltext{1}
{Joint ALMA Observatory, Alonso de C\'ordova 3107, Vitacura, Santiago, Chile}

% ESO Chile
\altaffiltext{2}
{European Southern Observatory, Alonso de C\'ordova 3107, Vitacura, Santiago, Chile}

% NRAO Cville
\altaffiltext{3}
{National Radio Astronomy Observatory, 520 Edgemont Rd, Charlottesville, VA, 22903, USA}

% NRAO Socorro
\altaffiltext{4}
{National Radio Astronomy Observatory, P.O. Box O, Socorro, NM 87801, USA}

% ESO Garching
\altaffiltext{5}
{European Southern Observatory, Karl-Schwarzschild-Str. 2, D-85748 Garching bei M\"unchen, Germany}

% IRAM
\altaffiltext{6}
{IRAM, 300 rue de la piscine 38400 St Martin d'H\`eres, France}

% ASIAA
\altaffiltext{7}
{Institute of Astronomy and Astrophysics, Academia Sinica, P.O. Box 23-141, Taipei 106, Taiwan}

% NAOJ
\altaffiltext{8}
{National Astronomical Observatory of Japan, 2-21-1 Osawa, Mitaka, Tokyo 181-8588, Japan}

% JAXA/ISAS
\altaffiltext{9}
{Institute of Space and Astronautical Science (ISAS), Japan Aerospace Exploration Agency (JAXA), 3-1-1 Yoshinodai, Chuo-ku, Sagamihara, Kanagawa 252-5210 Japan}

% Cambridge
\altaffiltext{10}
{Astrophysics Group, Cavendish Laboratory, JJ Thomson Avenue, Cambridge, CB3 0HE, UK}

% Bologna
\altaffiltext{11}
{INAF, Istituto di Radioastronomia, via P. Gobetti 101, 40129 Bologna, Italy}

% Robert Lucas
\altaffiltext{12}
{Institut de Plan\'etologie et d'Astrophysique de Grenoble (UMR 5274), BP 53, 38041, Grenoble Cedex 9, France}
	
% Manchester
\altaffiltext{13}
{Jodrell Bank Centre for Astrophysics, School of Physics and Astronomy, University of Manchester, Oxford, Road, Manchester M13 9PL, UK}

% KASI
\altaffiltext{14}
{Korea Astronomy and Space Science Institute, Daedeokdae-ro 776, Yuseong-gu, Daejeon 305-349, Korea}

% Leiden
\altaffiltext{15}
{Leiden Observatory, Leiden University, P.O. Box 9513, 2300 RA Leiden, The Netherlands}

% NRC NAASC
\altaffiltext{16}
{National Research Council Herzberg Astronomy \& Astrophysics, 5071 West Saanich Road, Victoria, BC V9E 2E7, Canada}

% Bonn
\altaffiltext{17}
{Argelander-Institute for Astronomy, Auf dem Hugel 71, D-53121 Bonn, Germany}

\email{cvlahaki@alma.cl}

\begin{abstract}
% Max 250 words.
We present initial results of very high resolution Atacama Large Millimeter/submillimeter Array (ALMA) observations of 
the $z$=3.042 gravitationally lensed submillimeter galaxy \hbox{HATLAS J090311.6+003906} (SDP.81). 
These observations were carried out using a very extended configuration as part of Science Verification for the 2014 ALMA Long Baseline Campaign, with baselines of up to $\sim$15~km. 
We present continuum imaging at 151, 236 and 290~GHz, at unprecedented angular resolutions as fine as 23~milliarcseconds (mas), corresponding to an un-magnified spatial scale of $\sim$180~pc at $z$=3.042. 
The ALMA images clearly show two main gravitational arc components of an Einstein ring, with emission tracing a radius of 
$\sim$1.5\arcsec. 
We also present imaging of \coten, \jeight, \jfive\/ and \watertrans\/ line emission. 
The CO emission, at an angular resolution of $\sim$170~mas, is found to broadly trace the gravitational arc structures but with differing morphologies between the CO transitions and compared to the dust continuum. 
Our detection of \water\/ line emission, using only the shortest baselines, provides the most resolved detection to date of thermal \water\/ emission in an extragalactic source. 
The ALMA continuum and spectral line fluxes are consistent with previous Plateau de Bure Interferometer and Submillimeter Array observations 
despite the impressive increase in angular resolution.
Finally, we detect weak unresolved continuum emission from a position that is spatially coincident with the center 
of the lens, with a spectral index that is consistent with emission from the core of the foreground lensing galaxy. 
\end{abstract}


\section{Introduction}\label{sec:intro}

\begin{figure*}
\includegraphics[width=\textwidth]{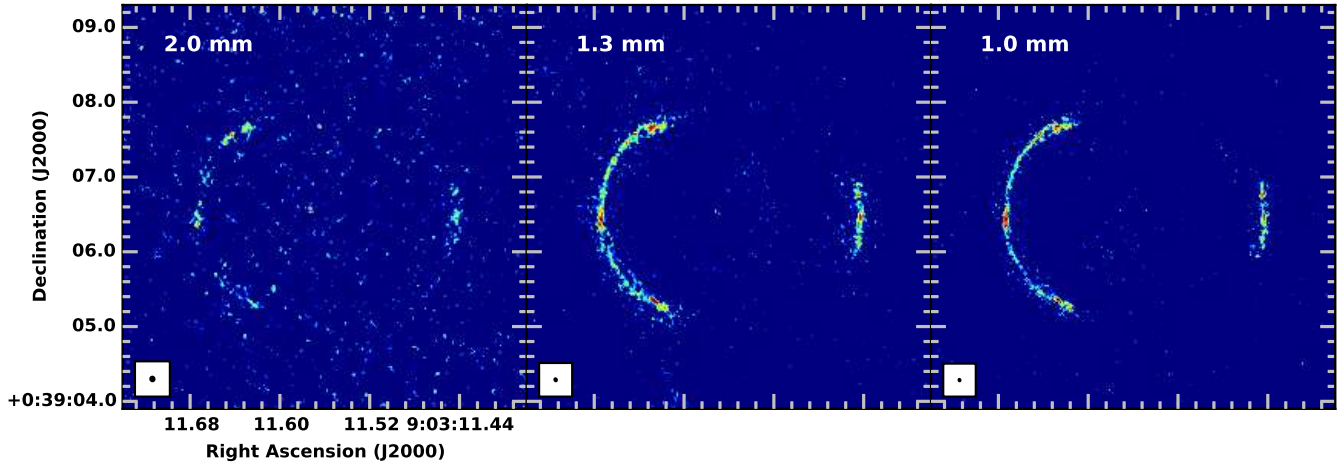}
\caption{High resolution ALMA images of SDP.81: 2.0 mm (Band 4), 1.3 mm (Band 6) and 1.0 mm (Band 7) continuum. The smallest synthesized beam is 31$\times$23~mas, for the Band 7 data. See Table~\ref{tab:cont} for other beam sizes. In addition to the gravitational arc structures, weak unresolved emission is also detected at a position that is coincident with the center of the foreground lensing galaxy.}\label{fig:highres}
\end{figure*}

Dusty star-forming galaxies make up a significant fraction of the star formation rate (SFR) space density of the Universe at z$\sim$2-3, 
where they are most numerous \citep[e.g.][]{Chapman2005,Casey2014}. 
A powerful probe of their properties is provided by gravitational lensing, with the magnification of both flux and apparent solid angle making it possible to study in detail regions of luminosity-redshift space that would otherwise be inaccessible. 
Surveys that cover large areas of sky, e.g. the {\it Herschel} Astrophysical Terahertz Large 
Area Survey \citep[H-ATLAS;][]{Eales2010}, {\it Herschel} HERMES \citep{Oliver2012} and 
the South Pole Telescope Survey \citep{Vieira2010}, have recently uncovered a large population of strongly-lensed sources, confirming previous predictions \citep{Blain1996,Negrello2007}. 
Other surveys, e.g. the Herschel Lensing Survey \citep[][]{Egami2010} are searching for lensed sources in a targeted way. 
A review is given by \citet{Lutz2014}.  

At submillimeter (submm) wavelengths, the study of red-shifted atomic and molecular lines in gravitationally lensed sources provides a useful diagnostic tracer of their interstellar medium (ISM) and star formation properties, active galactic nuclei (AGN), structure and dynamics \citep[e.g.][]{Swinbank2010,Swinbank2011,Riechers2011a, Fu2013}. 
In addition to CO lines, the magnification provided by lensing also makes it possible to observe molecular species that 
would otherwise be difficult to observe at high redshift, such as HCO$^{+}$, HCN or \water\/ lines \citep[e.g.][hereafter OM13]{Gao2007,Riechers2011b,Spilker2014,Omont2013}. \water\/ emission lines provide some of the strongest molecular lines in local Ultra-luminous Infrared Galaxies (ULIRGs) and starburst galaxies \citep[e.g.][]{GA2004,vdW2010}, and strong water lines have now been detected in a number of high-$z$ galaxies (OM13 and references therein).

HATLAS J090311.6+003906 (hereafter SDP.81) is a gravitationally lensed submm galaxy (SMG) at $z$=3.042 that was detected in H-ATLAS by \citet[][hereafter NE10]{Negrello2010}. 
Based on spectroscopic redshifts from observations of several CO lines, NE10 determined 
that SDP.81 is a background source at $z$=3.042 that is being lensed by a foreground elliptical galaxy at $z$=0.299.
Lens modeling analysis by \citet{Dye2014} and \citet[][hereafter BU13]{Bussmann2013} 
shows a magnification factor $\mu\sim$11. From NE14, the unmagnified infrared luminosity for the 
background source is 5.1$\times$10$^{12}$\,L$_{\odot}$ and the estimated intrinsic SFR is 527~M$_{\odot}$yr$^{-1}$.
SDP.81 has been studied in dust continuum emission using the Submillimeter Array (SMA) (e.g.~NE10) and in molecular lines (CO and \watertrans) 
using the IRAM Plateau de Bure interferometer (PdBI; OM13), achieving angular resolutions of $\sim$0.6\arcsec--3\arcsec. 
At this resolution, the dust continuum and molecular emission begin to 
resolve into multiple components, with the SMA and PdBI images indicating two gravitational arcs.  

Recently, the Atacama Large Millimeter/submillimeter Array (ALMA) carried out unprecedented high angular resolution continuum and spectral line observations of SDP.81 as part of 
Science Verification (SV) of its 2014 Long Baseline Campaign. 
Details of the campaign, whose aim was to demonstrate the scientific capability of ALMA to observe on baselines of $\sim$10~km or more, are described in an accompanying paper \citep[][hereafter AL15]{Alma2015a}. 
In this paper, we present the multi-wavelength SDP.81 ALMA SV observations.
Section~\ref{sec:obs} describes the observations and data reduction. In Section~\ref{sec:res}, we present the data and 
initial results, and in Section~\ref{sec:disc}, 
we discuss initial conclusions and the potential these rich data offer for detailed future analysis.

Throughout this paper, we adopt a cosmology with $H_{0}$=67~\kms\,Mpc$^{-1}$, $\Omega_{m}$=0.32, $\Omega_{\Lambda}$=0.68 \citep{Planck2014}.

\section{Observations and Data Reduction}\label{sec:obs}

Observations of SDP.81 in continuum, CO lines and an \water\/ line were carried out in 2014 October as part of the 2014 ALMA Long Baseline Campaign 
using ALMA's most extended configuration to date. 
These data are publicly available from the ALMA Science Portal\footnote{http://www.almascience.org}. 
The ALMA array consisted of 23 nominal long baseline antennas in a preliminary configuration, plus a number of other antennas at shorter baselines (see AL15). 
During the periods of observation, between 22-36 antennas were in the array (with the number of available antennas varying with observing band) and baseline lengths ranged from $\sim$15~m to 15~km. Only 10\% of the baselines were shorter than ~200~m. 
Observations were carried out in Band 4 ($\sim$2~mm), Band 6 ($\sim$1.3~mm) and Band 7 ($\sim$1.0~mm). 
Initial test observations in Band 3 ($\sim$3~mm) proved unfruitful, primarily due to the faintness of the continuum 
in this band due to the steep falloff of the thermal dust spectrum. 

For each of Bands 4, 6, and 7, the total available 7.5~GHz bandwidth was divided into four spectral windows (spws): three 15.6~MHz channel width spws for the 2.0, 1.3, and 1.0~mm continuum, respectively, and one spw for the targeted spectral line. 
For Band 4, a 0.976~MHz channel width spw was centered on the redshifted \cofive\/ line  
($v_{\rm rest}$ = 576.267~GHz); for Band 6, a 0.976~MHz or 1.95~MHz channel width spw (online spectral averaging was applied for some executions) was centered on the redshifted low-excitation water line \watertrans\/ ($v_{\rm rest}$ = 987.927 GHz ($E_{\rm up}$ = 101 K)); for Band 7, a 1.95~MHz channel width spw covered the redshifted \coten\/ line ($v_{\rm rest}$ = 1151.985~GHz). 
In Band 6, one of the spectral windows included the \coeight\/ line ($v_{\rm rest}$ = 921.799~GHz).
Observed frequencies are given in Tables~\ref{tab:cont} and~\ref{tab:line}. Other details of the observations, tunings and correlator setup are given in AL15 and Appendix~\ref{details}.

$Uv$-coverage was critically important for imaging this target, due to both its complex morphology and its
location near the celestial equator that resulted in non-optimal horizontal 
$uv$-tracks\footnote{http://casaguides.nrao.edu/index.php?title=ALMA2014\_LBC\_SVDATA}. 
A relatively large amount of  total observing time ($\sim$9-12 hours per band, or 9-12 executions of the scheduling block) was therefore necessary to achieve the good hour angle coverage needed for good image quality. The total on-source integration times were 5.9, 4.4 and 5.6 hours in Bands 4, 6, and 7, respectively.

The data were reduced using the Common Astronomy Software Application package
\citep[{\textsc CASA\footnote{http://casa.nrao.edu}};][]{mcmullin2007}. The procedure is described in the scripts available on the ALMA 
Science Portal\footnote{http://www.almascience.org}.  
Imaging was carried out using a {\it robust}=1 weighting of the visibilities. Given the preliminary antenna configuration, there was a lack of $uv$-coverage for 200-500~m baselines, and thus the use of robust weighting ({\it robust}=1) was critical for achieving acceptable image quality (see Appendix~\ref{details} and AL15 for further details). 
None of the data were self-calibrated, since our attempts to perform self-calibration showed that there was insufficient signal-to-noise on the longest baseline antennas.
The resulting synthesized beam sizes range from $\sim$23--60~mas.

For the spectral lines, it was necessary to taper the CO line $uv$ data to 1000 k$\lambda$ in order to achieve good detections, resulting in coarser angular resolutions of $\sim$170~mas. 
For the \water\/ data, it was necessary to taper the $uv$ data further, to 200 k$\lambda$ ($\sim$0.9\arcsec\/ resolution). 
The CO data were binned spectrally into channels 21~\kms\/ wide. For the narrower \water\/ line, we binned the data into 10.5~\kms\/ channels. All the spectral line data were continuum subtracted and were imaged using rest frequencies corresponding to $z$=3.042. 

The resulting synthesized beam sizes and rms noise levels for the continuum and spectral line images are given in Tables~\ref{tab:cont} and~\ref{tab:line}, respectively. 

\begin{figure*}
\includegraphics[width=\textwidth]{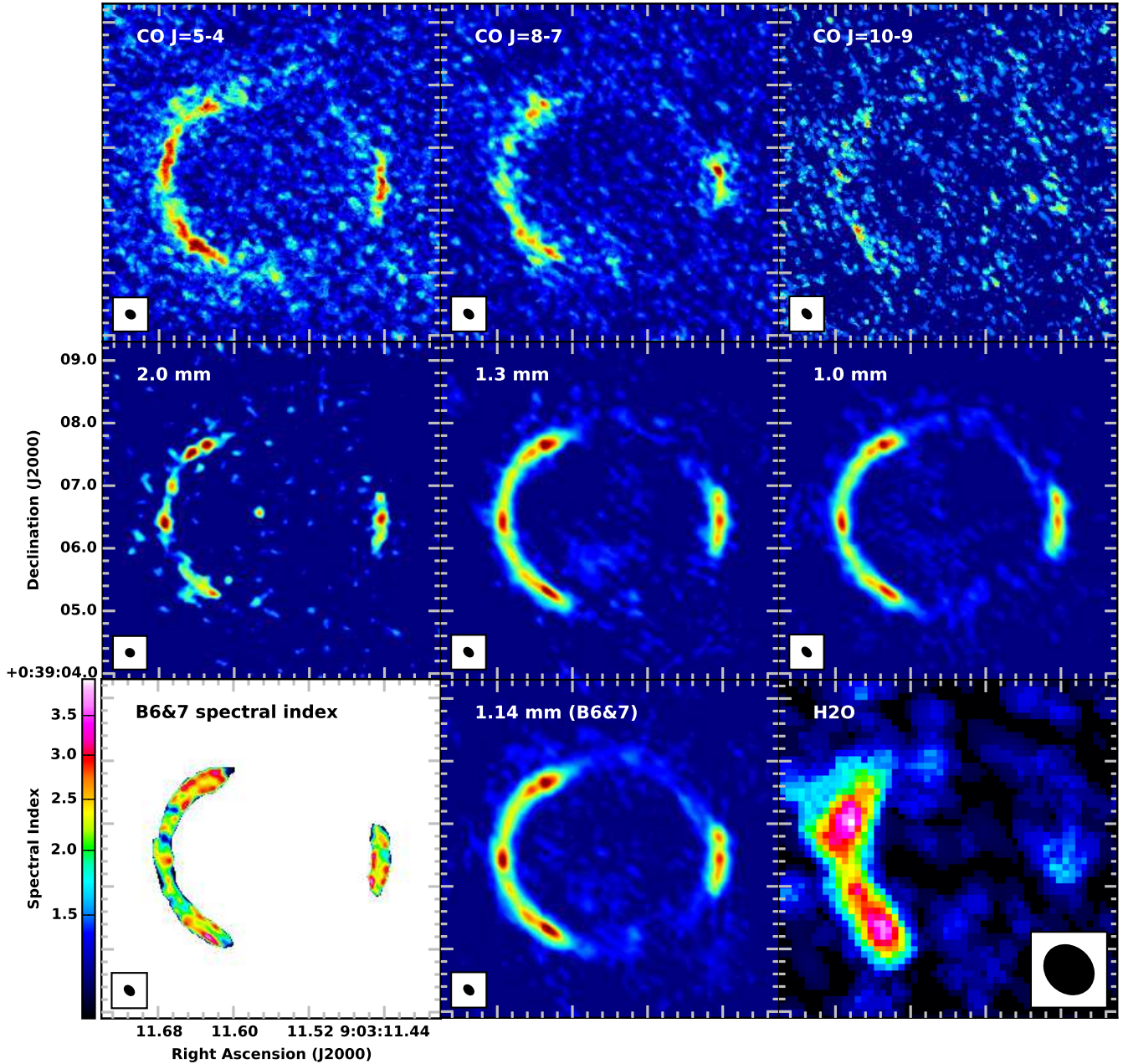}
\caption{ALMA images with $uv$-tapering to 1000~k$\lambda$ (CO lines and continuum) or 200~k$\lambda$ (H$_{2}$O line). Top: CO~$J$=5-4, 8-7 and 10-9 velocity integrated intensity. Middle: 2.0 mm, 1.3 mm, and 1.0 mm continuum. Bottom: Band 6\&7 spectral index, 1.14mm continuum (combined Band 6 \& 7 data; see Appendix~\ref{details}), and H$_{2}$O velocity integrated intensity. Beam sizes are $\sim$170~mas, except for \water\/ which has a larger $\sim$0.9\arcsec\/ beam (see Table~\ref{tab:line} for spectral line details).}\label{fig:taperres}
\end{figure*}

\section{Results}\label{sec:res}

\begin{deluxetable*}{ccccccccccc}
\centering
\tabletypesize{\scriptsize}
%\rotate
\tablecaption{ALMA continuum parameters for SDP.81\label{tab:cont}}
\tablewidth{0pt}
\tablehead{
\colhead{Band} & Frequency & \colhead{Component} & \colhead{$S_{\nu}\,pk$} & \colhead{$\sigma_{rms}$} &  \colhead{$\theta_{b}$} & \colhead{$S_{\nu}$}  & \colhead{$\sigma_{rms}$} & \colhead{$\theta_{taper}$}\\
     &      \colhead{(GHz)}     &  & \colhead{[$\mu$Jy~beam$^{-1}$]} & \colhead{[$\mu$~Jy]} & \colhead{[mas (deg)]} &  \colhead{[mJy]} & \colhead{[$\mu$~Jy]} & \colhead{[mas (deg)]}
}
\startdata
   4  & 151  & Total                & ...                           & 8  &  56$\times$50 (18)    & 3.5$\pm$0.1 & 11  & 134$\times$118  (61) \\
 &   (2.0~mm)      &       E        &   520$\pm$ 8   &  &                            & 2.8$\pm$0.1       &   & \\
                &      &     W        &   353$\pm$8     &   &                           & 0.7$\pm$0.1       &   &  \\
   6    & 236  & Total       & ...                             & 10  & 39$\times$30 (20)  &   26.3$\pm$1.3  &  21  &  164$\times$114 (47)  \\
   & (1.3~mm)     &       E       &   141$\pm$10     &     &                         &   19.3$\pm$1.3  &  &  \\
                  &   &     W        &   112$\pm$10     &      &                        &   7.0$\pm$0.4     & &   \\
    7  & 290   &  Total     &   ...                             & 9  &  31$\times$23 (16) &  37.7$\pm$1.5     & 21  & 170$\times$106 (43) \\
 &   (1.0~mm)     &     E       &    124$\pm$9    &       &                        &  27.7$\pm$1.0    & &      \\
               &     &   W         &   112$\pm$9      &        &                      &  10.0$\pm$0.2    & &    \\
6\&7   & 262 & Total        & ...                           &   ...    &      ...                    &   32.9$\pm$1.0          & 15  & 163$\times$110 (44)\\
 &  (1.14~mm)    &      E      &    806$\pm$15   &  ...     &     ...                     &   23.5$\pm$0.5      &   &  \\
                &   &   W       &   627$\pm$15      &   ...   &          ...          &   7.0 $\pm$0.2    &    &   \\
\enddata
\tablecomments{
Column (3) Spatial components: eastern arc (E), western arc (W) and the total over all the emission. 
(4) Peak flux density in the specified arc component. Values were measured in the high resolution images whose beam sizes, $\theta_{b}$, are listed in column (6), except for the combined Band 6\&7 data for which we used the $uv$-tapered image (see Section~\ref{sec:obs}). Uncertainties are 1$\sigma$ where $\sigma$ is the RMS noise level in the high resolution images, given in (5). For Band 6\&7 the 1$\sigma$ uncertainty is the RMS noise value given in (8). 
(7) The integrated flux density over the E and W arcs separately, and the total emission over all the arc components (see Section~\ref{sec:cont}). Values are measured in the tapered images, whose beam sizes, $\theta_{taper}$, are given in column (9). Uncertainties were calculated from the larger of either the rms variation of the flux density in apertures placed at source-free locations in the images or [no. independent beams]$^{0.5}\times$3$\sigma$ where $\sigma$ is the RMS noise level given in (8). Values do not include 5\% absolute flux calibration uncertainty.
(8) RMS noise level in the $uv$-tapered images.
Synthesized beam sizes of the high resolution and tapered images, and the beam position angle, are given in (6) and (9), respectively.}
\end{deluxetable*}

\begin{table}
\begin{center}
\caption{ALMA flux densities at the position of the foreground lens\label{tab:lens}}
\begin{tabular}{ccc}
\tableline\tableline
Band & Frequency &  $S_\nu pk$\\
 & (GHz) & ($\mu$Jy~beam$^{-1}$) \\
\tableline
4 & 151 (2.0~mm)  & 60$\pm$8  \\
6 & 236 (1.3~mm) & 50$\pm$10  \\
7 & 290 (1.0~mm) & 43$\pm$9 \\
\tableline
\end{tabular}\\
\tablecomments{Frequency is the observed frequency. Peak fluxes were measured in the high resolution images whose beam sizes are given in Table~\ref{tab:cont}. Uncertainties are 1$\sigma$ where $\sigma$ is the rms noise level given in Table~\ref{tab:cont} and do not include 5\% absolute flux calibration uncertainty.}
\end{center}
\end{table}

\begin{table*}
\begin{center}
\caption{ALMA CO and H$_{2}$O line parameters for SDP.81\label{tab:line}}
\begin{tabular}{clccccccccc}
\tableline\tableline
Band & Line & $\nu_{obs}$  & Component  &  $I_{l} pk$ & $I_{l}$ & $\Delta V_{l}$ & $V$& $\mu$L$_{l}$ & $\mu$L$'_{l}$$$/10$^{10}$  & $\theta_{taper}$\\
         &   &  (GHz)  &      & (mJy~beam$^{-1}$) & (\jykms)& (\kms) & (\kms) & (10$^{8}$ L$_{\odot}$) &  (K~\kms\,pc$^{2}$) & [mas (deg)] \\
\tableline
4     &  \cofive\/ & 142.570   & Total (fit) &  ...   &  9.2$\pm$1.2& ... & ... & 9.6$\pm$0.3 & 15.6$\pm$0.5  & 155$\times$121 (57)\\
%       & &   &Total   & & (7.2)  & ... & & ... \\
   & & & R & 26.5$\pm$0.2 & 6.5$\pm$0.3 & 231$\pm$6 &72$\pm$5 & 6.8$\pm$0.3 &  11.0$\pm$0.5 &... \\
   & & & B & 9.0$\pm$0.2 & 2.7$\pm$0.3 & 282$\pm$40 & $-$174$\pm$19 & 2.8$\pm$0.3 & 4.6$\pm$0.5 & ...\\
      &  & & E      & ...            & 7.2$\pm$0.2  & ...& ... & ...&...&... \\
      & & & W     &...            & 1.2$\pm$0.2  &...  & ... & ...&...&...\\

6     &  \coeight\/ & 228.055 & Total (fit) & ...  &   9.3$\pm$0.4& ... & ... & 15.5$\pm$0.7 & 6.2$\pm$0.3 & 169 $\times$ 117 (47)\\
       & &&R   &   28.0$\pm$0.2 & 6.6$\pm$0.4 & 221$\pm$11& 77$\pm$6 & 11.0$\pm$0.7 & 4.4$\pm$0.3 & ...\\
        & &&B   & 10.2$\pm$0.2   & 2.7$\pm$0.4 & 249$\pm$44 & $-$204$\pm$15 & 4.5$\pm$0.7 & 1.8$\pm$0.3 &...\\
%       &&Total & ...  & ...& 12.0 &  &... \\
      & &&E     & ...            &  7.2$\pm$0.4 &  ... &... &...& ...&... \\
      &&& W    & ...            &  2.0$\pm$0.1 & ... &... &... &...&... \\

       & H$_{\rm 2}$O$~(2_{\rm 02}-1_{\rm 11}$)  & 244.415 &  Total (fit)    & 9.4$\pm$0.9 &  2.0$\pm$0.1  & 196$\pm$15  & 76$\pm$7 & 3.6$\pm$0.2 & 1.2$\pm$0.1 & 852$\times$721 (45) \\
%         &   & TOT       & ... &   & ...  &  &  ...\\
         &   & & E      & ...  & 1.6$\pm$0.1 & ... & ... & ...& ...& ... \\
         &   && W      & ...  & 0.15$\pm$0.06  &... & ... & ...&...& ... \\

7 & \coten\/& 285.004 & Total (fit) & ... & 3.2$\pm$0.2 & ... & ... & 6.7$\pm$0.1 & 1.4$\pm$0.1 & 172$\times$112 (42)\\
&  && R & 10.1$\pm$0.3 & 1.8$\pm$0.2 & 170$\pm$18 & 75$\pm$8 & 3.7$\pm$0.1 & 0.8$\pm$0.1 & ...\\
 & && B & 6.7$\pm$0.3 & 1.3$\pm$0.2 & 188$\pm$30 & $-$204$\pm$12 & 2.7$\pm$0.1 & 0.6$\pm$0.1 & ...\\
%       & \coten\/ &Total  & ...    &  ... &  & &... \\
    &  &&E              & ... & 1.2$\pm$0.1 & ... & ... & ...& ...& ... \\
       & && W              & ...  & 0.4$\pm$0.1 &  ... &... & ...& ... & ...\\
\tableline
\end{tabular}\\
%\tablenotetext{}{}
\tablecomments{Parameters and errors for the CO and H$_{2}$O lines are derived from Gaussian fits (Figure~\ref{fig:spectra}) for the two spectral components R and B (Section~\ref{sec:co}). 
All values are derived from the $uv$-tapered images, whose beam sizes and position angles are listed in the final column.
$\nu_{obs}$ is the observed frequency at $z$=3.042, which corresponds to zero velocity offset.  
$I_{l}$ is the velocity-integrated intensity of the CO or H$_{2}$O line and $\Delta V_{l}$ is the line width (FWHM).
Luminosities, L$_{l}$, are calculated following \citet{Solomon1992}. 
$\mu$ is the magnification factor (Section~\ref{sec:intro}). 
Values of the integrated intensity for the two main spatial components (E and W arcs; Section~\ref{sec:cont}) are also given, measured directly on the integrated intensity images (Section~\ref{sec:spectral}). Uncertainties on $I_{l} pk$ are 1$\sigma$, where $\sigma$ is the rms noise level per channel. Rms noise levels for \cofive, \coeight\/ and \coten\/ are  0.20, 0.15 and 0.25~mJy per 21~\kms channel, respectively. For \water, the rms noise level is 0.9 ~mJy per 10.5~\kms channel. Error values do not include 5\% absolute flux calibration uncertainty.}
\end{center}
\end{table*}

\subsection{Dust continuum emission}\label{sec:cont}

In Figure~\ref{fig:highres} we present the high resolution continuum images at 151, 236 and 290~GHz. 
The highest angular resolution achieved, 23~mas, is an impressive increase of a factor of 
$\sim$20--80 in angular resolution compared to previous SMA and PdBI observations of this source and corresponds to an unmagnified spatial scale of $\sim$180~pc at $z$=3.042 (or of the order of a few tens of parsecs in the source plane). 
Continuum images tapered to a similar resolution as the spectral line images ($\sim$170~mas) are presented in Figure~\ref{fig:taperres}.

Two main gravitational arc components of an Einstein ring are observed in all images: a larger eastern arc (E) that contains the highest peak intensity,  and a smaller western arc (W). Several distinct surface brightness peaks are observed in each arc.
The morphology is consistent with previous observations at similar wavelengths with the PdBI and SMA (OM13, NE10).
The locations of the arc components in the highest resolution (236 and 290~GHz) images 
lie on a ring of $\sim$1.55\arcsec\/ radius that is consistent with values for the Einstein radius $\theta_{E}$=1.52\arcsec--1.62\arcsec\/ derived by NE10, BU13 and \citet{Dye2014}  from SMA 880~\micron\/ or HST data. 
Measured ALMA flux densities, both spatially integrated over all the arc components and for the major arc components (E and W) separately, are given in Table~\ref{tab:cont}. These were computed within apertures that matched the spatial extent of the continuum emission in each band using the tapered images. The tapered Band 6, Band 7 and combined Band 6\&7 continuum images also show a suggestion of low S/N features that appear to trace out a more complete ring in addition to the two main arcs; due to the low significance of these features we did not include them in our flux measurement apertures. 
Peak fluxes were measured from the high resolution images. 
The ALMA 2.0~mm (236~GHz) continuum flux density is in agreement with the PdBI 244~GHz value (27$\pm$1.2~mJy; OM13) within the uncertainties. 
The 1.0, 1.14, 1.3 and 2.0~mm flux densities are consistent with the SED of the background source presented by NE14. 

%\bigskip

\subsubsection{Detection of continuum emission at the central position of the foreground lens}
In all three bands, we have detected continuum emission located at the central position of the foreground lens (SDSS~J090311.57+003906.5). 
This is visible in all the images in Figure~\ref{fig:highres} and the 2~mm continuum tapered image in Figure~\ref{fig:taperres}. 
From the 236~GHz image, the centroid of the emission peak is 
at 09$^{h}$03$^{m}$11$^{s}$.57 $+$00${\degr}$39${\arcmin}$06${\arcsec}$.53 (J2000). 
The angular distance from this position to the eastern arc is 1.52$\pm$0.02\arcsec, which is consistent with the $\theta_{E}$ values determined by previous authors. 
The emission is unresolved, even at 290~GHz, where the beam size is only 31$\times$23~mas. 
 At the redshift of the lens ($z$=0.299; NE10), 23~mas corresponds to a spatial scale of $\sim$100 pc.
 The detection of this central emission suggests two main possibilities: a previously undetected AGN in the lens or 
an additional image of the background source that is predicted to arise close to the center of the lensing galaxy \citep[e.g.][and references therein]{Hezaveh2015}. 
The ALMA continuum flux values at this positon differ by orders of magnitude from the SED predictions of both the lens and source in NE14, and we also note that we do not detect the central source in the CO or \water\/ line data. 
Although the elliptical galaxy is massive (NE10 find a mass within the Einstein radius of log(M$_{E}$)=11.56~M$_{\odot}$), we find no evidence in the literature of any AGN activity. 
However, the spectral index we derive from the ALMA 151, 236 and 290~GHz flux densities in Table~\ref{tab:lens} and the 1.4~GHz FIRST flux density (0.61~mJy) is $-$0.49$\pm$0.13, which implies that the majority of the emission we have detected originates from the core of the foreground elliptical galaxy.

\subsubsection{Spectral index}
The combined Band 6 and Band 7 spectral index image, obtained from data tapered to 1000 k$\lambda$ (see Section~\ref{sec:obs}), 
is shown in Figure~\ref{fig:taperres}. Pixels $<$4$\sigma$ have been masked.
The mean spectral index in the unmasked regions is 2.34$\pm$0.61, with the mean value measured in the W and E arcs comparable within the uncertainties (2.45$\pm$0.72 compared to 2.30$\pm$0.57). The range of values is consistent with 
dust spectral indices from 1.4-4. We note, however, that from the flux densities given in Table~\ref{tab:cont}, which represent the total emission region as opposed to the higher S/N regions defined by the spectral index image, the average spectral index is 1.8$\pm$0.2.

\subsection{CO $J$=10-9, $J$=8-7 and $J$=5-4 line emission}\label{sec:co}

\subsubsection{Properties of the CO images}\label{sec:spectral}

CO~\jten, \jeight\/ and \jfive\/ velocity integrated intensity images are presented in Figure~\ref{fig:taperres}. 
The CO morphology is broadly consistent with the overall two-arc morphology seen in continuum, 
with a larger and higher surface brightness E arc and a smaller and generally lower surface brightness W arc.  
The symmetric distribution of the continuum peaks, however, is not matched in CO, with the
CO emission appearing more clumpy throughout the arc structures. The CO emission also appears to trace a
somewhat more extended and less well-defined ring than the continuum at the same angular resolution.
For \cofive, the peak integrated intensities are comparable in both arcs and the morphology bears more similarity to the continuum.
The peak \coten\/ and \coeight\/ intensities, however,  are respectively observed in the E and W arcs, and the \coten\/ emission exhibits a different spatial morphology altogether, with several compact regions of emission along the arc structures. This may be due to differences in the spatial origin of the higher-$J$ CO emission, or simply due to the lower S/N in the \coten\/ data. 
Notably, the spatial locations of peak intensity are not coincident between the CO transitions.

\begin{figure}\label{fig:spec}
%\vspace{1cm}
\centering
\includegraphics[width=0.42\textwidth, angle=270]{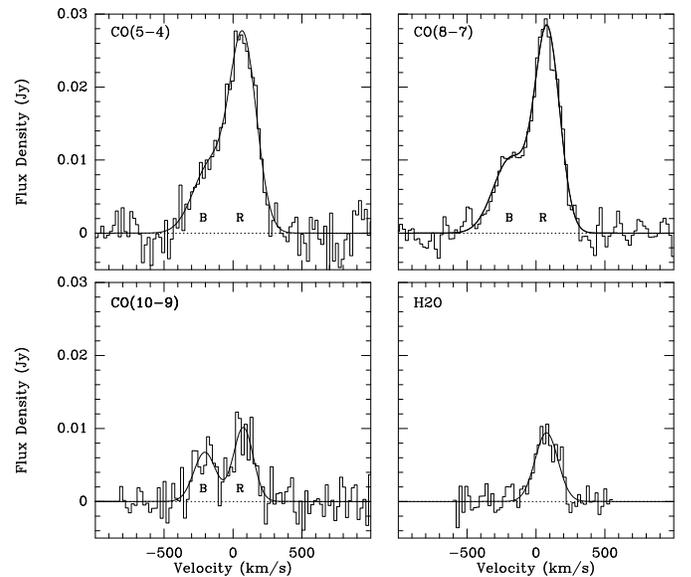}
\caption{ALMA CO $J$=5-4, CO $J$=8-7, CO $J$=10-9 and \watertrans\/ spectra for SDP.81. Spectra were spatially integrated over regions containing the E and W arc components (except for \water, for which we only included E; see Section~\ref{sec:cont}) in the $uv$-tapered images and the resulting spectra fitted with Gaussians. The two, redshifted and blue-shifted, components found for the CO transitions are marked with `R' and `B', respectively.}\label{fig:spectra}
\end{figure}

Fitted spectral profiles, spatially integrated over the two arc components, are given in Figure~\ref{fig:spectra}. The spectra were produced by spatially integrating the CO emission over a polygon region encompassing the E and W arcs as defined on CO integrated intensity images produced from all pixels with a S/N$>$3 in the channels containing emission. 
All parameters were derived from images that had first been smoothed to the same resolution (0.2\arcsec). The CO parameters derived from Gaussian fits are given in Table~\ref{tab:line}. For comparison, integrated intensity values for the E and W components are also given, and were measured directly on the integrated intensity images. For all measurements we do not include any low level emission outside the main arc components (see Section~\ref{sec:cont}). 
The ALMA \cofive\/ integrated intensity measured over both the arc components, peak intensity and spectral shape are consistent with the PdBI \cofive\/ results presented by OM13 (the PdBI integrated intensity is 7.0$\pm$0.4~Jy~\kms). 
The three CO transitions have asymmetrical line profiles (Figure~\ref{fig:spectra}), 
with a red-shifted component (R) that is stronger than the blue-shifted component (B), 
which is again consistent with OM13. 
The B component is relatively brighter than R in the higher-$J$ CO lines, suggesting that the B emission originates
from hotter gas than the R component. 
The line widths are in good agreement with
 the CO~$J=$1-0  and \cofive\/ values found by \citet{Frayer2011} and OM13, and are within the range found for other typical SMGs \citep[e.g.][]{Greve2005}. 

\begin{figure*}
\centering
\includegraphics[width=0.8\textwidth]{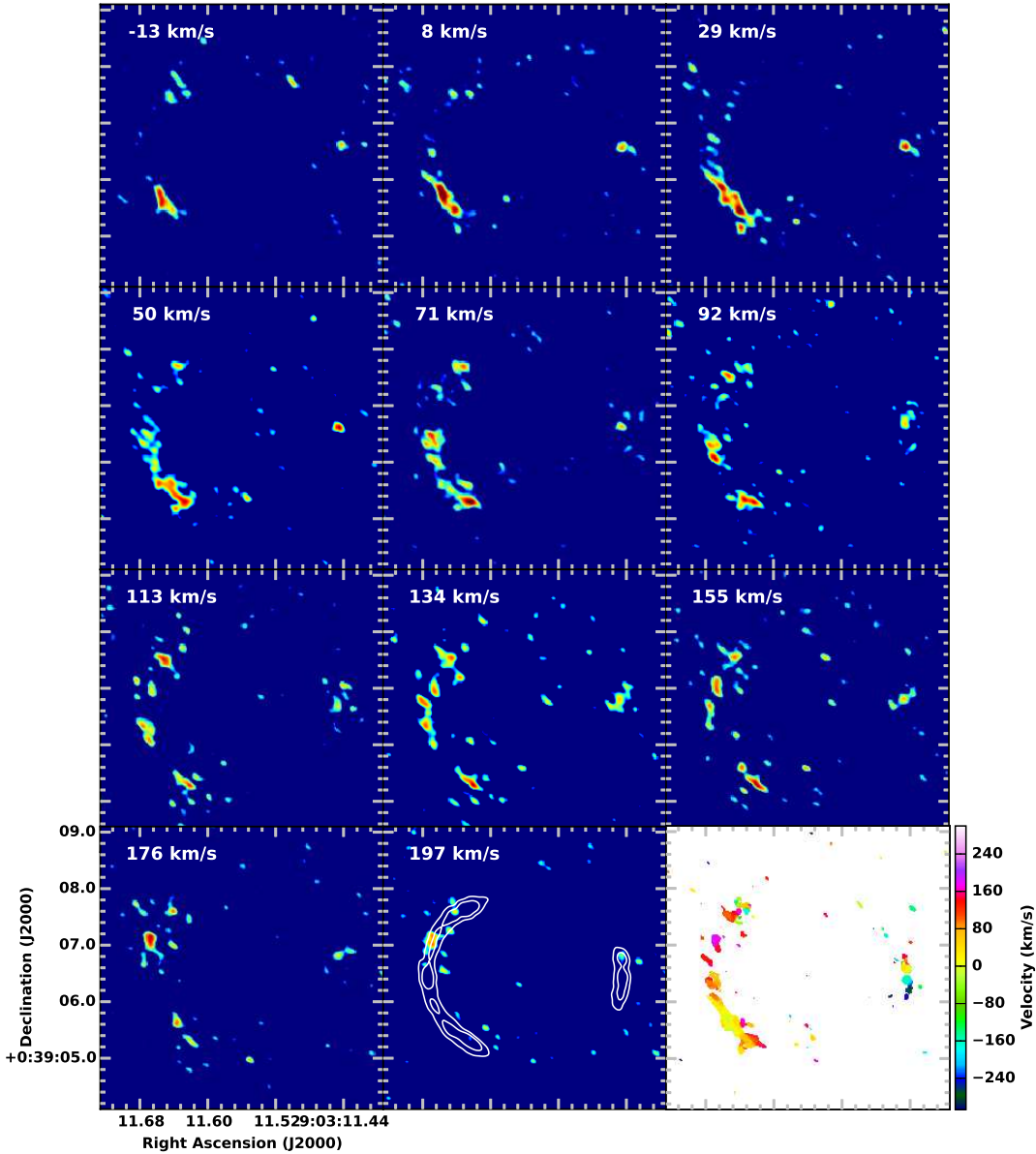}
\caption{Channel maps for the CO~$J$=8-7 emission, showing the red-shifted component, R. The R component was defined from $-$55 to 239~\kms\/ based on the spectral profiles in Figure~\ref{fig:spectra}. Note that for clarity of the figure we do not plot the three outermost channel ranges. 
As a guide, contours of the combined Band 6\&7 continuum emission are shown in one panel, at 0.2 and 0.4 mJy~beam$^{-1}$ (15 and 30$\sigma$). Last panel: \coeight\/ velocity field, for both B and R components.}\label{fig:chanmap}
\end{figure*}

\begin{figure*}
\centering
\includegraphics[width=0.65\textwidth]{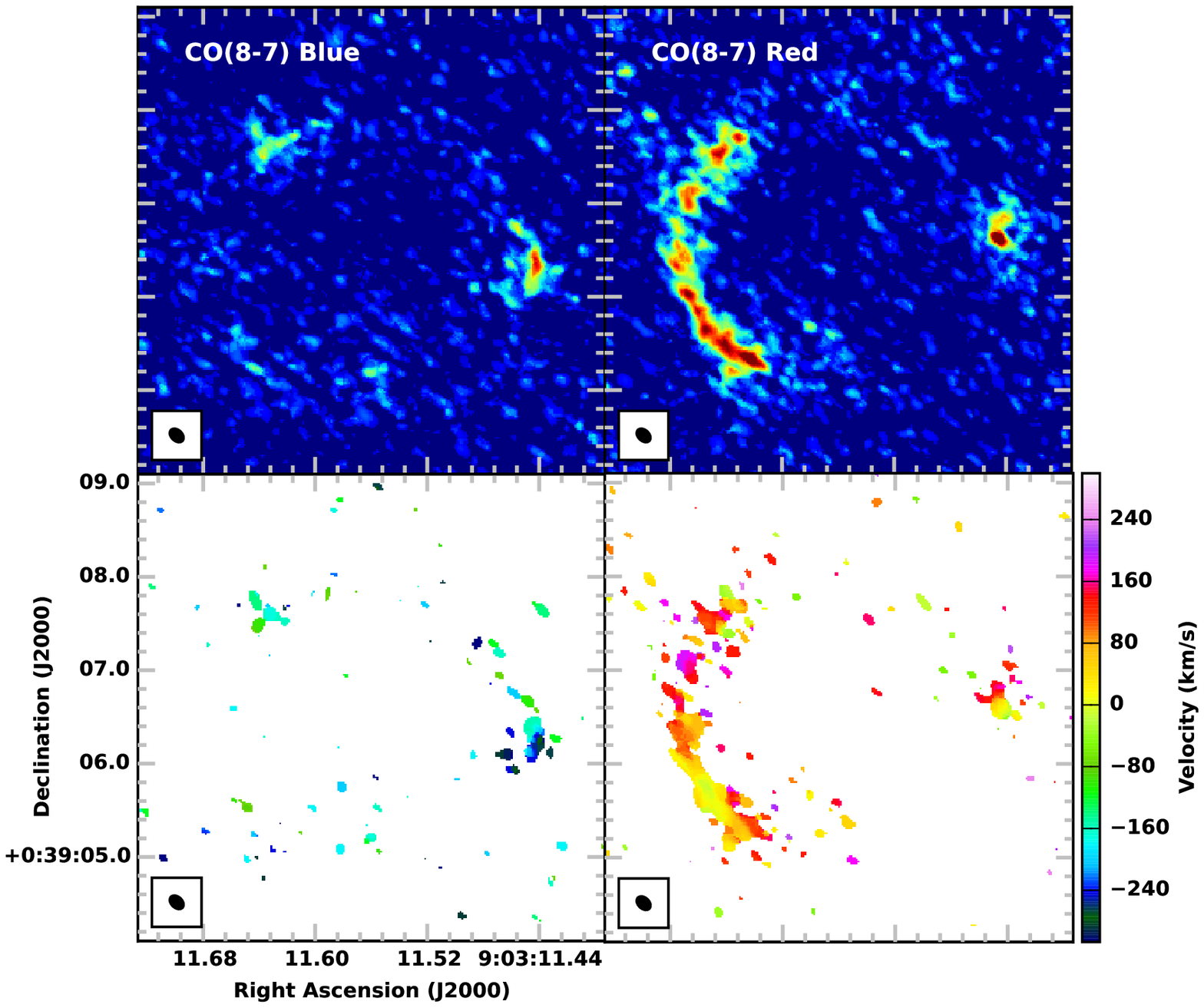}
\caption{CO~$J$=8-7 integrated intensity (top) and velocity field (bottom) images for the B and R components. 
The B and R components were defined from $-$307 to -76~\kms\/ and $-$55 to 239~\kms, respectively, based on their spectral profiles (Figure~\ref{fig:spectra}).}\label{fig:rb}
\vspace{3ex}
\end{figure*}

\subsubsection{Velocity structure}

In Figure~\ref{fig:chanmap}, we show the \coeight\/ channel maps corresponding to the R component seen in Figure~\ref{fig:spectra}, in which a north-south velocity gradient is observed for the E arc.
We do not show channel maps for the B component since organized velocity structures are not clearly discernible in B.
Similar properties are seen for the \cofive\/ emission. 
This velocity structure of R is also demonstrated by \coeight\/ integrated intensity and velocity field images produced separately for the R and B components (Figure~\ref{fig:rb}). Here, in addition to the N-S velocity gradient seen for R, it is also clear that the R and B image components have different spatial locations in the image plane.
Spatially, the E arc corresponds predominantly to the R spectral 
component, with only a small part of the E arc that is blue-shifted. The emission in the W arc, conversely, 
has a symmetrical profile with roughly equal contributions from the B and R components. 
 A velocity gradient is apparent for the W arc that encompasses both the B and R components, which can be seen in the velocity field image in Figure~\ref{fig:chanmap}. 
The presence of two distinct spectral components, and the observed velocity structure in Figures~\ref{fig:chanmap} and~\ref{fig:rb}, 
is suggestive of complexity in the source structure, which has been previously suggested from  lens modelling analysis (e.g. BU13).

\subsubsection{CO line ratios}\label{sec:coratios}

From the CO line luminosities given in Table~\ref{tab:line}, we find CO line brightness temperature ratios of $r_{85}$=0.5$\pm$0.1, $r_{108}$=0.2$\pm$0.1 and $r_{105}$=0.1$\pm$0.1. 
Taking the CO~$J$=1-0 value of 1.1~Jy\,\kms\/ from \citet{Frayer2011} we also find a ratio $r_{51}$=0.3$\pm$0.1. This is consistent with the value $r_{31}$=0.5 reported by \citet{Frayer2011}. 
\citet{Lupu2012} detected high-$J$ CO lines in SDP.81 but some were blended with other lines so we do not include them here. 
The values of the line ratios suggest the presence of a low-excitation gas component, which is consistent with results found for other SMGs \citep[e.g.][]{Carilli2010,Danielson2011,Hodge2013,Bothwell2013} and the suggestion that most SMGs contain a significant proportion of cool, moderate-density, extended gas \citep[e.g.][]{Ivison2011}. 
Taking the total IR luminosity $\mu$L$_{IR}=$5.4$\times$10$^{13}$\,L$_{\odot}$ (uncorrected for magnification) from NE14, the ratio L$_{IR}$/L$'_{CO}$ is 346$\pm$62~L$_{\odot}$(K~\kms\/pc$^{2})^{-1}$ for \cofive, which is towards the upper end of the range for local ULIRGS \citep[][]{Solomon1997} and in good agreement with the value found for typical SMGs \citep[e.g.][(360$\pm$140~L$_{\odot}$(K~\kms\/pc$^{2})^{-1}$)]{Greve2005}.

\subsection{Water emission}\label{sec:water}
The \watertrans\/ image is shown in Figure~\ref{fig:taperres}. 
Emission is clearly detected in the E arc and exhibits a morphology that appears consistent with those seen in continuum and CO. 
Despite the tapering required to achieve a good detection, this is the highest resolution detection of thermal \water\/ emission in this source to date. 
For the W arc, there is a suggestion of possible emission, with integrated and peak intensities that are $\sim$2$\sigma$ in the integrated intensity image. The velocity range associated with this feature ($\sim$200~\kms) is consistent with the B component. Due to the low significance of the W arc detection we do not include it in our spectrally fitted measurement. 
The ALMA integrated intensity and spectral profile are in good agreement with the PdBI value (1.8$\pm$0.5~Jy~\kms). The spectrum (Figure~\ref{fig:spectra}) shows that the \water\/ emission originates from the R component. The central velocity of the \water\/ line profile agrees with all the CO line profiles within the uncertainties of the fits. 

\bigskip

\subsubsection{H$_{2}$O line ratios}

The \water/high-$J$ CO line ratio is widely considered a useful diagnostic of PDR versus XDR conditions \citep[e.g.][]{GA2010}. 
We find an \water/\coeight\/ line ratio, $I_{H_{2}O}$/$I_{CO}$ of 0.2$\pm$0.1. 
This is relatively low compared to results in the literature for other active galaxies. 
\citet{Omont2011} find a value of 0.5$\pm$0.2 for SDP.17b, while values of 0.4$\pm$0.2 and  0.75$\pm$0.23 have been found for the Cloverleaf and Mrk231 \citep[][]{Bradford2009,GA2010}. 
Taking a value of $L_{IR}$ as in Section~\ref{sec:coratios}, we find an L$_{H2O}$/L$_{IR}$ ratio of 6.7$\times$10$^{-6}$, which is similar to the value found for Arp~220 (OM13 and references therein). Both the intrinsic values of L$_{H2O}$ and L$_{IR}$ are at the lower end of the range for other high-$z$ sources in the sample of OM13.

\section{Discussion}\label{sec:disc}
The results presented here on SDP.81 provide a first view of the wealth of information contained in these ALMA data. Providing a leap in angular resolution compared to previous observations, these data demonstrate the power of ALMA to image gravitationally lensed systems with high resolution and high fidelity.
 
With the continuum images, we have presented strong detections of thermal dust emission from the $z$=3.042 background source at unprecedented angular resolutions, with beam sizes as small as 23~mas corresponding to spatial scales in the source on the order of an impressive few tens of parsecs. At this high angular resolution we clearly detect the two main arc structures of the Einstein ring, but using only the shorter baselines we find evidence of low-level emission that traces a more complete ring.

We have also detected weak ($\sim$5$\sigma$) continuum emission that is spatially coincident with the center of the foreground lens, and find a 1.4--290~GHz spectral index of $\sim$$-$0.5 that implies we have detected emission from an AGN in the foreground elliptical galaxy that must have a very low accretion rate.

Our detection of \watertrans\/ in SDP.81, at 0.9\arcsec\/ resolution, is the highest resolution detection to date of thermal water emission in an extragalactic source. 
\water\/ emission from the W component is tentatively detected for the first time, with a contribution to the total \water\/ emission of only a few per cent.

We have also presented detections of three transitions of CO, \jfive, \jeight, and \jten, at 170~mas resolution, and shown that while they spectrally have similar red- and blue-shifted components, their spatial morphologies are rather different.
The properties of the CO and \water\/ line data indicate that SDP.81 may be a complex source, as was suggested
by previous authors \citep[e.g. BU13,][]{Dye2014}.

The overall similarity in the spatial and spectral distributions of the CO and \water\/ lines suggests that both species may have a common origin, although the sub-thermal ($\ll$1) CO line ratios and strong detections of high-$J$ CO lines could suggest the presence of an extended cool and moderate-density gas component \citep[e.g.][]{Harris2010}. 
Our CO line measurements suggest that the higher-$J$ CO lines are less luminous, which may indicate that any AGN present would be subdominant, 
and thus that a complex mixture of PDRs and XDRs would be required to adequately describe the excitation in SDP.81.
The \water/CO(8-7) ratio is lower in SDP.81 than for other high-$z$ sources in the literature \citep[e.g.][OM13]{Omont2011}, which suggests that either we are missing significant extended \water\/ emission or that the \water\/ and CO excitation is relatively low. The former does not seem likely, given the similarity between the ALMA and PdBI results. 
It is possible that differential lensing \citep[e.g.][]{Blain1999,Serjeant2012} is playing a role in the relatively weak \water\/ emission in the W component. 
Without detailed modeling, any effect from differential magnification cannot be quantified, and in a highly magnified system such as SDP.81, differential lensing could potentially lead to, for example, changes in the spectral indices, the CO ladder or line ratios.

Future studies will be able to quantify this effect in SDP.81 through detailed foreground mass modelling \citep[][]{Serjeant2012,Bourne2014}, and the combination of this high angular resolution ALMA data with near-IR data will provide constraints on the differential lensing between the stellar and dust emission \citep[e.g. ][]{Calanog2014}. 
Detailed lens modelling is beyond the scope of this work, but will be crucial for understanding the full nature of these ALMA observations.

\acknowledgments

This paper makes use of the following ALMA data: \dataset{ADS/JAO.ALMA\#2011.0.00016.SV}. 
ALMA is a partnership of ESO (representing its member states), NSF (USA) and NINS (Japan), together with NRC (Canada), 
NSC and ASIAA (Taiwan), and KASI (Republic of Korea), in cooperation with the Republic of Chile. 
The Joint ALMA Observatory is operated by ESO, AUI/NRAO and NAOJ.
The National Radio Astronomy Observatory is a facility of the National 
Science Foundation operated under cooperative agreement by Associated 
Universities, Inc. 
This research made use of Astropy, a community-developed core Python package for Astronomy \citep{Astropy2013}. 
We thank Alastair Edge for helpful comments.

{\it Facilities:} \facility{ALMA}.

\appendix

\section{Data reduction details}\label{details}

The phase center was 09$^{h}$03$^{m}$11$^{s}$.61 $+$00${\degr}$39${\arcmin}$06${\arcsec}$.7 (J2000). Precipitable water vapor (PWV) values at zenith varied in the ranges 0.6-3.2, 0.5-3.1 and 0.3-0.7~mm for Bands 4, 6, and 7, respectively, which is better than average weather conditions for the given Band. 
The Band 6 and Band 7 data were corrected with the most accurate antenna pad positions that were measured by the end of the 
campaign (see AL15).
The phase calibrator, J0909+0121, was $\sim$1.65\degr\/ away from the target and was typically observed every 78~s for all three bands. 
Bandpass and flux calibration were performed using observations of the quasars 
J0825+0309 and J0854+2006 (or in a few cases J0750+1231 or J1058+0133), respectively,  in each execution of the 
scheduling block (SB).  For each dataset, between one and four antennas were flagged during the calibration process.

Continuum images were integrated over a bandwidth of $\sim$7~GHz and produced using multi-frequency synthesis, a {\it robust}=1 weighting of the visibilities, and {\it nterms}=1. The use of robust weighting ({\it robust}=1) was found to be critical for achieving acceptable image quality given the lack of $uv$-coverage for 200-500~m baselines due to the preliminary antenna configuration (since a limited number of antenna relocations were possible during the period of the campaign; AL15). 
The total number of antennas in the array depended on observing band, with fewest antennas in Band 4, due to fewer available Band 4 receivers on antennas on the shortest baselines. 
{\it Multi-scale} imaging \citep{Cornwell2008} was used, 
with scales of 0, 5 and 15 for both the continuum and spectral line imaging (0, 5, 15 and 45 in Band 4). As the source lies sufficiently close to the phase center, the images have not been corrected for the effect of primary beam attenuation. This effect will be $<$5\% in all bands, and thus smaller than the overall flux calibration uncertainty.

We experimented with imaging the combined Band 6 and Band 7 $uv$ data with {\textsc CLEAN} mode $nterms$=2, which attempts to account for the spectral index of emission across the observed frequency band \citep{Rau+Cornwell2011}. The SNR of the full-resolution $uv$ data was found to be too low for the algorithm to determine the spectral index, except around the strong peaks of emission, which resulted in a flux scale in the combined image that was not reliable. We were, however, able to make a combined (1.14~mm) image using Band 6\&7 continuum images that were $uv$-tapered to match the resolution of the spectral line data, which yielded a robust spectral index image at $\sim$170~mas resolution.

\end{document}